\documentclass[journal=nalefd,manuscript=letter]{achemso}

\usepackage{chemformula} 
\usepackage[T1]{fontenc} 
\usepackage{comment}
\usepackage{graphicx}
\usepackage{hyperref}
\usepackage{amssymb}
\usepackage{derivative}
\usepackage{caption}
\usepackage{upgreek}
\usepackage{dirtytalk}

\author{Archit Negi}
\affiliation[kyudai]{Department of Physics, Kyushu University, Nishi-ku, Fukuoka 819-0395, Japan}
\alsoaffiliation[kyodai]
{Department of Chemical Engineering, Kyoto University, Nishikyo-ku, Kyoto 615-8246, Japan}
\author{Ryota Sakamoto}
\affiliation[yale]{Department of Biomedical Engineering, Yale University, 10 Hillhouse Avenue, New Haven, CT, USA}
\author{Makito Miyazaki}
\affiliation[riken]{RIKEN Center for Integrative Medical Sciences, 1-7-22 Suehiro-cho, Tsurumi-ku, Yokohama, Kanagawa 230-0045, Japan}
\alsoaffiliation[riken2]{RIKEN Center for Biosystems Dynamics Research, 2-2-3 Minatojima-minamimachi, Chuo-ku, Kobe, Hyogo 650-0047, Japan}
\alsoaffiliation[riken3]{Graduate School of Medicine, Science and Technology, Shinshu University, 3-1-1 Asahi, Matsumoto, Nagano 390-8621, Japan}
\author{Yusuke T. Maeda}
\affiliation[kyodai]{Department of Chemical Engineering, Kyoto University, Nishikyo-ku, Kyoto 615-8246, Japan}
\email{maeda@cheme.kyoto-u.ac.jp}

\title[An \textsf{achemso} demo]
{Myosin-driven advection and actin reorganization control the geometry of confined actomyosin gel}

\abbreviations{IR,NMR,UV}

\begin{document}







\begin{abstract}
Harnessing nanoscale motor proteins to actively control material shape is a promising strategy in nanotechnology and material science. One notable system is the actomyosin network, composed of actin filaments and myosin motor proteins, providing a natural platform for constructing contractile, shape-adaptive materials. While the role of actomyosin in shaping cells has been extensively studied, the reverse question - how boundary shape affects the actomyosin system - remains poorly understood. Here, we present a microfabricated system that reveals how geometrical confinement directs the organization of actomyosin networks within microwells. By combining experimental and numerical analysis, we show that the asymmetric shape of the microwells is transferred to contracted actomyosin gels via myosin-driven actin flow. Furthermore, tuning myosin contractility and actin polymerization rate allows control over the size and shape of actomyosin gels. Our findings provide a bottom-up framework for integrating molecular motors and cytoskeletons into confined architectures to create responsive biomaterials.
\end{abstract}


\section{Introduction}
Living cells come in a wide variety of shapes, and the shape of a cell is often linked to its function \cite{murrell2015forcing}. Inside cells, the nucleus or organelles have definite shapes that are tightly regulated, and abnormalities of their shapes often reflect deficiencies in underlying signaling and force regulation that can significantly impair cellular functions and even cause diseases \cite{organelle_disease_1, organelle_disease_3}. However, this does not mean that intracellular structures are static; they can change in response to mechanical cues from the environment to fulfill the functional needs of the cell \cite{vogel2006local}. An important contributor to this regulation is the active cytoskeleton, driven by autonomous molecular motor activity \cite{banerjee2020actin}. Having the ability to precisely control the shape of biomaterials has the potential for many applications, including in biotechnology and nanotechnology, such as the fabrication of functional hydrogels \cite{guan2021bio}.

One of the major active cytoskeletons is actomyosin, a network-structured complex of actin filaments and myosin molecular motor protein. The actomyosin network plays a central role in numerous processes, which often involve deformation of the cell membrane \cite{murrell2015forcing}. For example, during cell migration, actin and myosin self-organize to form an actomyosin cortex that generates cortical tension \cite{paluch2006dynamic}, while at the last step of cell division, actomyosin forms a contractile ring and separates one mother cell into two daughter cells \cite{kruse2024actomyosin}. Such diverse self-organized dynamics are driven by myosin motor proteins that generate forces by converting chemical energy into mechanical work \cite{sakamoto2024mechanical}. Previous studies have designed the spatial pattern of myosin generated forces, controlling the transiently contracting actomyosin network \cite{schuppler2016boundaries}. However, the dynamic assembly and disassembly of actin filaments can also alter their spatiotemporal distribution within a cellular space \cite{actin_dynamics}. Thus, it is becoming increasingly apparent that cell-sized confinement is essential to create dynamic behaviors such as periodic actin flow and reorganization of actomyosin with sustained contraction \cite{miyazaki2015cell, Sakamoto2020, sakamoto2022geometric, PhysRevResearch.5.013208, fakhri_sciadv, keren_eLife, Malik-Garbi2019, malik_elife, Krishna2024}. Hence, how the geometry of cellular spaces leads to various macroscopic self-organizations of the actin cytoskeleton, linked to nanoscale forces exerted by myosin, is not fully understood.

In this study, we use a reconstituted actomyosin system to demonstrate how the shape of cell-like boundaries regulates the self-organization of the actomyosin network. We design the spatial confinement geometry using microfabrication to define the mechanical contribution of the boundary shape. Our integrated experimental and numerical analysis reveals that the asymmetric shape of the confinement is transferred to the shape of contracted actomyosin gels formed inside the microwells through actin flow driven by nanosized molecular motors. Moreover, we demonstrate that the shape and size of the actomyosin gel can be regulated by dynamic variables, such as myosin contractile force and actin polymerization rate, thereby establishing design principles for the morphology of this system. 



\section{Results}
We used cytoplasmic extracts from the eggs of \textit{Xenopus laevis} frogs, which consist primarily of the actin cytoskeletal proteins (filamentous actin or F-actin, and actin monomer), the Arp2/3 protein complex, which nucleates actin polymerization, and the myosin molecular motor protein, which binds to F-actin and forms the actomyosin network \cite{mitchison_gelation,wuhr2014deep}. To visualize F-actin, we used tetramethylrhodamine (TMR)-labeled LifeAct, which binds to F-actin but not to monomeric actin \cite{lifeact2009}.

In bulk extracts, the F-actin network is polymerized and myosin generates contractile forces, resulting in a gelation-contraction process \cite{mitchison_gelation}. When confined to cell-sized spaces, such as inside a water-in-oil droplet with a phospholipid monolayer interface (hereinafter referred to as lipid droplet), the actomyosin network contracts to form a nucleus-like cluster of actomyosin gel, coupled with a continuous inward flow of actin \cite{zoher_actin_cur_bio, zoher_actin_pnas}. The confined actomyosin network system has been shown to position these clusters depending on the size of the confinement\cite{Malik-Garbi2019, malik_elife, Sakamoto2020, PhysRevResearch.5.013208, Krishna2024}. 

Here, to determine the effect of the confinement shape on the organization of the actomyosin network, we fabricated microwells using NOA81 (a UV-curable adhesive) passivated with PEG-PLL and confined the cell extracts inside them (Figure \ref{fig:fig_1}(a)). Using the microwells, we can precisely control the shape and size of the confinements, which is not possible in a lipid droplet \cite{yamamoto_noa, inoue2024surface, araki2021controlling, zaferani2024building, yamazaki2024controlling}. The shape of the microwells was that of a circular segment, defined by the length of the flat side \(W\) and the diameter \(D\), with the \(W/D\) ratio ranging from 0.0 (circle) to 1.0 (semicircle). The observations were made through confocal microscopy, the focus being near the top of the microwell (see Materials and Methods).

\begin{center}
\includegraphics[width=0.9\textwidth]{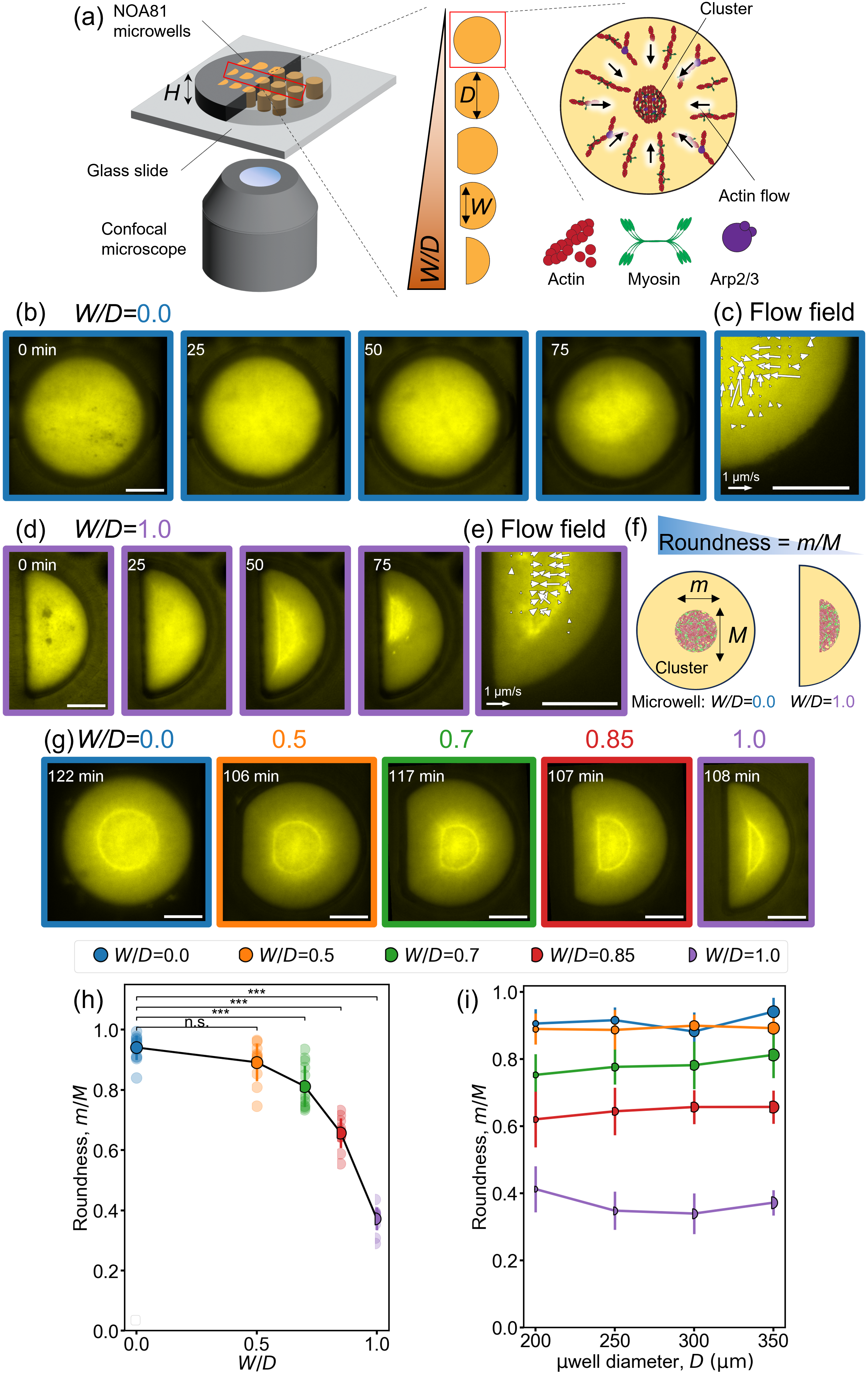}
\end{center}

\begin{figure}
\caption[]{\small{Confinement geometry determines the shape of the contractile actomyosin gel. (a) Schematic of the experiment. The cell extracts containing the actomyosin network were confined inside NOA81 microwells in the shape of circular segments and the dynamics were observed through confocal microscopy. (b) Timelapse of the actomyosin dynamics in a circular microwell (\(D=350\ \upmu\)m, \(W/D=0.0\)) showing the contraction of the actin network into a cluster; F-actin is visualized with TMR-labeled LifeAct. (c) Velocity field of the actin flow in a circular microwell, determined through particle image velocimetry. (d) Timelapse of the actomyosin dynamics in a semicircular microwell (\(D=350\ \upmu\)m, \(W/D=1.0\)) showing the actin network contracting to a non-circular cluster. (e) Velocity field in a semicircular microwell. (f) Schematic showing the dependence of cluster roundness on the microwell shape; \(m,\ M\) are respectively the minor and the major axes of the ellipse fitted to the cluster shape, and are used to define the roundness, \(R=m/ M\). (g) Variation of the cluster shape as the microwell \(W/D\) changes. (h) Dependence of cluster roundness on the microwell shape. (i) Dependence of cluster roundness on microwell diameter and shape. All microwells in (g), (h) are of diameter \(D=350\ \upmu\)m. At least \(n=11\) and \(n=4\) microwells were used for averaging at each data point in (h) and (i) respectively. The \(p\)-values were calculated using the Mann-Whitney \textit{U} test, *** means \(p < 0.001\), n.s. means \(p \geq 0.05\). Scale arrows, 1 \(\upmu\)m/s. Scale bars, 100 \(\upmu\)m.  }}
\label{fig:fig_1}
\end{figure}


To test whether the microwells show behavior similar to that of a lipid droplet, we first confined the cell extracts inside a circular microwell with diameter \(D=350\ \upmu\)m and height \( H \approx 150\ \upmu\)m. We found that the actomyosin network underwent an initial contraction and soon periodic actin flows appeared from the boundary (Figure \ref{fig:fig_1}(c) and S1, Movie S1). The contracted actomyosin then began to accumulate at the center of the microwell and eventually we observed that a circular cluster had formed (Figure \ref{fig:fig_1}(b)). Along the Z axis, the cluster was localized close to the top of the microwell (Figure S1). In the XY plane, it was observed to be close to the microwell center in 83\(\%\) of the circular microwells (Figure S2), consistent with the findings in lipid droplets of similar sizes \cite{Sakamoto2020}. From the microwell timelapse kymograph (Figure S1), the time period and velocity of the actin waves were determined to be \( T \approx 102\ \)s and \( v \approx 1.27\ \upmu\)m/s respectively, these values being comparable to the lipid droplet case \cite{Sakamoto2020}.

Next, to examine the effect of the confinement shape on cluster formation, we confined the cell extracts inside a semicircular microwell \(( W/D = 1.0 )\) with the same diameter \(D=350\ \upmu \)m and height \( H \approx 150\ \upmu \)m. We found that, as in the circular case, actin flows also appeared in this asymmetric microwell (Figure \ref{fig:fig_1}(d), \ref{fig:fig_1}(e) and S3, Movie S2). Actin accumulation and cluster formation also occurred inside this microwell, but interestingly, the cluster shape was not circular (Figure \ref{fig:fig_1}(d)). This suggests that the asymmetric shape of the microwell was transferred to the contracted actomyosin gel (Figure \ref{fig:fig_1}(f)). In the remaining microwells, we found that the cluster shape changed from a circle to a crescent as the microwell shape changed from circular to semicircular (Figure \ref{fig:fig_1}(g)).

To quantify the asymmetry of the cluster shape, we defined the roundness parameter \(R\), calculated by fitting an ellipse with a major axis \(M\) and a minor axis \(m\) to the cluster shape and taking the ratio \(R\)=\(m/M\). The closer the roundness is to \(R\)=1, the more circular the shape becomes. Figure \ref{fig:fig_1}(h) plots how roundness varies with the microwell shape, and we see a clear transition of the cluster shape from a circular to a non-circular geometry as the microwell shape changes from a circle \((W/D=0.0)\) to a semicircle \((W/D=1.0)\). Next, we checked whether the cluster shape would change depending on the size of the microwells. We found that the cluster roundness did not show a noticeable dependence on the diameter of the microwell at any \(W/D\). In contrast, roundness decreased as \(W/D\) increased from \(0\) to \(1\), regardless of microwell diameter (Figure \ref{fig:fig_1}(i)).

\begin{figure}
\includegraphics[width=\textwidth]{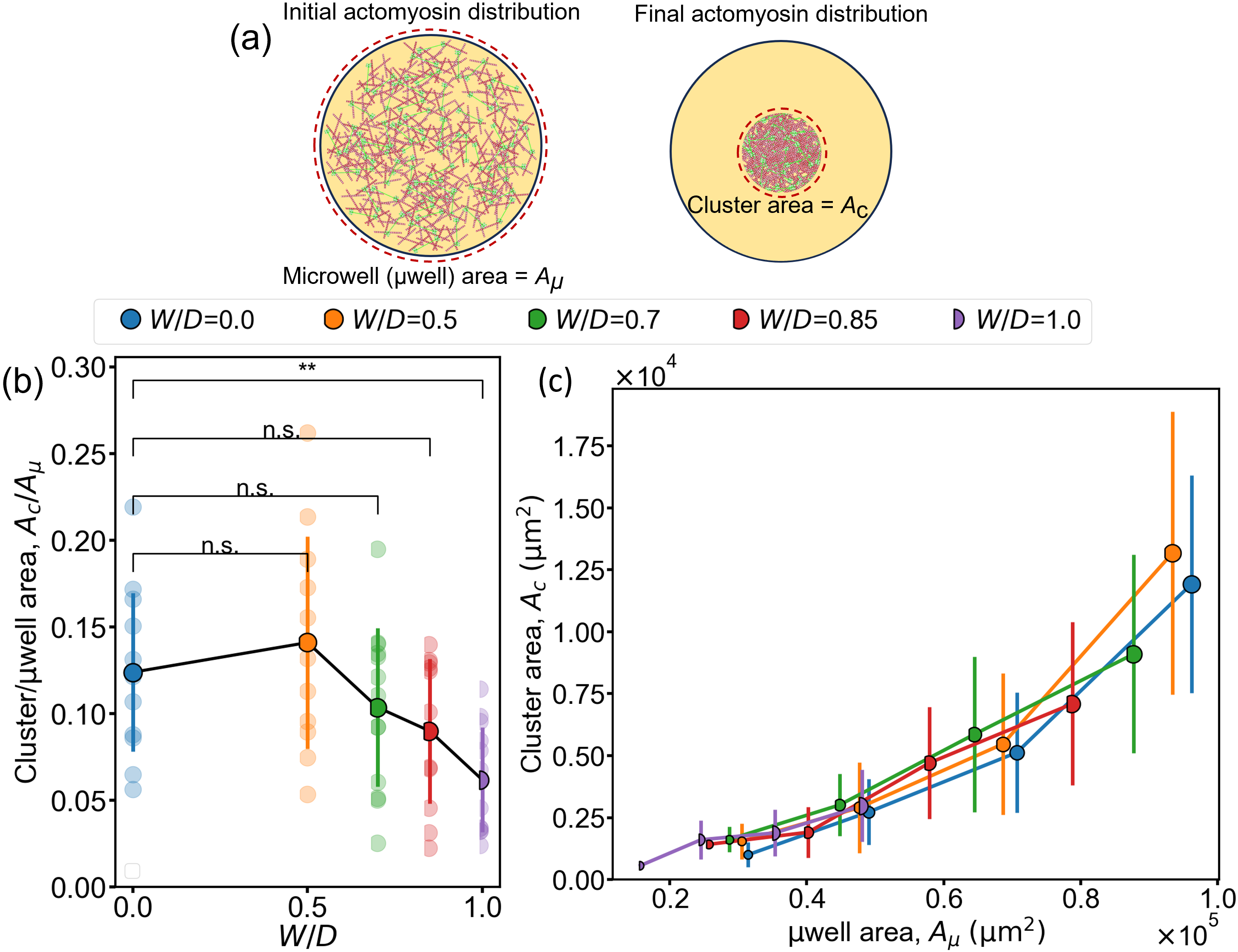}
\caption{\small{Confinement geometry determines the compressibility of actomyosin gel. (a) Schematic showing the contraction of the actomyosin network to form a cluster inside the microwell. (b) Dependence of cluster to microwell area ratio \((A_{\mathrm{c}} / A_{\mu})\) on microwell shape. (c) The cluster area increases as the microwell area increases and microwells with different shapes but similar sizes result in clusters of similar areas. At least \(n=11\) and \(n=4\) microwells were used for averaging at each data point in (b) and (c) respectively. The \(p\)-values were calculated using the Mann-Whitney \textit{U} test, ** means \(p < 0.01\), n.s. means \(p \geq 0.05\).}}
\label{fig:fig_2}
\end{figure}

We also evaluated the variation in the size of the clusters (Figure \ref{fig:fig_2}(a)), which was taken to be its cross-sectional area in the XY plane. For microwells with \(D=350\ \upmu \)m, we found that the size of the clusters decreased as the microwell shape changed from circular to semicircular, but the change in the ratio of cluster area to microwell area was not statistically significant, except when comparing the circular and semicircular microwells (Figure \ref{fig:fig_2}(b)). Upon comparing microwells with different diameters, we found that the cluster area increased with increasing microwell area (Figure \ref{fig:fig_2}(c)). Notably, the cluster area was comparable for microwells of different shapes but having similar areas, suggesting that the cluster size depends on the total amount of actin protein present in the system \cite{Sakamoto2020}.

\begin{figure}
\includegraphics[width=0.9\textwidth]{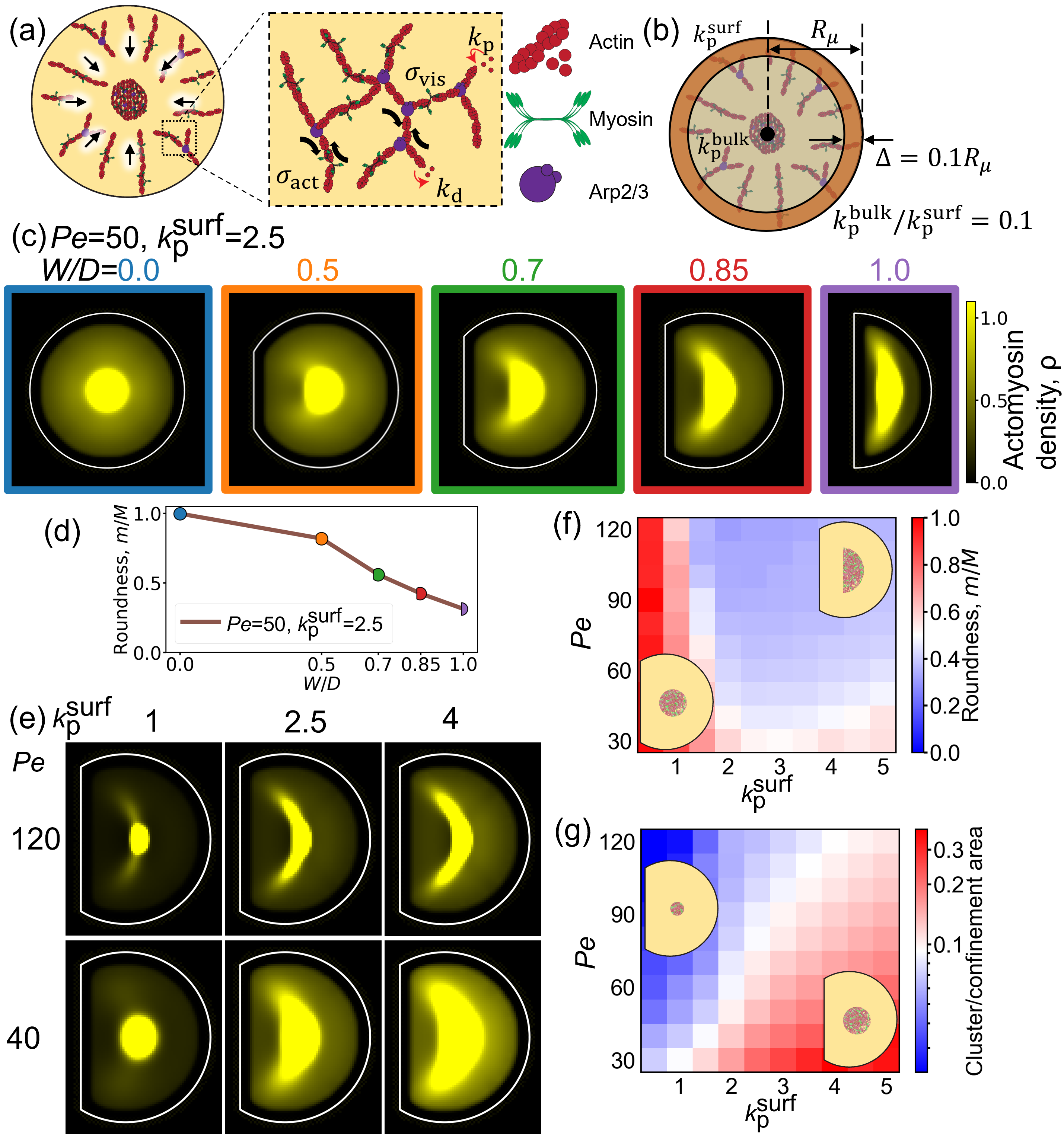}
\caption{\small{Active fluid model reproduces the shape and compressibility. (a) Schematic showing the parameters of the theoretical model. (b) Schematic showing the relation between the surface \( (k_{\mathrm{p}}^{\mathrm{surf}}) \) and the bulk \( (k_{\mathrm{p}}^{\mathrm{bulk}}) \) polymerization rates. (c) Numerical simulations of our theoretical model (\(Pe=50,\ k_{\mathrm{p}}^{\mathrm{surf}}=2.5\))  show the cluster shape to be dependent on the confinement shape, in qualitative agreement with the experiments. (d) Dependence of cluster roundness on the confinement shape for the simulations in (c). (e) Simulations show that the shape and size of the cluster changes with the P\'{e}clet number \(Pe\) and polymerization rate \(k_{\mathrm{p}}^{\mathrm{surf}}\). (f) Phase diagram showing the dependence of cluster shape on \(Pe\) and \(k_{\mathrm{p}}^{\mathrm{surf}}\); roundness decreases as \(k_{\mathrm{p}}^{\mathrm{surf}}\) increases. (g) Phase diagram showing the dependence of cluster size on \(Pe\) and \(k_{\mathrm{p}}^{\mathrm{surf}}\); area decreases as \(Pe\) increases while it increases as \(k_{\mathrm{p}}^{\mathrm{surf}}\) increases.}}
\label{fig:fig_3}
\end{figure}

To gain physical insights into how the confinement geometry regulates the final shape of the actomyosin gel, we used a numerical simulation model of an active fluid with spatial constraints \cite{kumar2014pulsatory, PhysRevResearch.5.013208} and modified it to account for asymmetric shapes by using the phase-field model. In this model, the actomyosin network is considered to be an active fluid, meaning that in addition to experiencing a passive viscous stress (\(\sigma_{\mathrm{vis}}\)), it also produces an active stress (\(\sigma_{\mathrm{act}}\)) due to the presence of myosin (Figure \ref{fig:fig_3}(a)). Myosin density is assumed to be proportional to F-actin density, simplifying the mass conservation to a single component. The dynamics of the system is given by considering force balance and mass conservation, with the nondimensionalized equations of motion being:
\begin{align}
\mathbf{v} &= \nabla^{2} \mathbf{v} + \lambda \nabla (\nabla \cdot \mathbf{v}) + \nabla f \label{vel}  \\
\pdv{\rho}{t} + Pe\nabla \cdot (\rho \mathbf{v}) &= \nabla^{2} \rho + k_{\mathrm{p}} - k_{\mathrm{d}} \rho + \epsilon_{\rho} \nabla^{2} \frac{\delta E}{\delta \rho} \label{mass} \\
\pdv{\phi}{t} &= D_{\phi} \nabla^{2} \phi + \Gamma_{\phi} U'(\phi) \label{phi}
\end{align}

Equation \eqref{vel} gives the velocity (\(\mathbf{v}\)) of the active fluid and is derived by equating the force due to the sum of active and viscous stresses with the friction force exerted by the confinement. Here \(f\) gives the density dependence of the active stress, and \(\lambda = 1 + \eta_{{b}} / \eta\), \(\eta_{{b}}\) and \(\eta\) being the bulk and dynamic viscosities, respectively. To simplify the momentum balance, we assume \(\lambda = 1/3\) \cite{LeGoff2020}. Equation \eqref{mass} describes the conservation of actomyosin mass and it is the diffusion advection equation for actomyosin density (\(\rho\)), which also has a source (actin polymerization, \(k_{\mathrm{p}}\)) and a sink (actin depolymerization, \(k_{\mathrm{d}}\)). \(Pe\) in this equation is the P\'{e}clet number, the ratio of advective and diffusive time scales, given by \(Pe=(\zeta \Delta \mu)_0 / D \gamma\), where \((\zeta \Delta \mu)_0 \) is contractile stress per F-actin due to myosin, \(D\) is the diffusion constant and \(\gamma\) is the friction coefficient. \(E\) is an effective energy that ensures a no-flux boundary condition. Equation \eqref{phi} is the phase-field equation that determines the boundary of the confinement based on the value of the phase field parameter (\(\phi\)). We also assume the polymerization rate of actin to be higher near the surface of the confinement \((k_{\mathrm{p}}^{\mathrm{surf}})\) compared to the bulk \((k_{\mathrm{p}}^{\mathrm{bulk}})\), with \(k_{\mathrm{p}}^{\mathrm{bulk}}/k_{\mathrm{p}}^{\mathrm{surf}}=0.1\) (Figure \ref{fig:fig_3}(b)), as polymerization is enhanced at the boundary in lipid droplets \cite{Sakamoto2020, zoher_actin_pnas}, a condition necessary for wave generation \cite{PhysRevResearch.5.013208}. This assumption holds for all of the following results.

On solving these equations numerically for a circular confinement, we found that a steady inward flow of the active fluid was observed, followed by the formation of a circular cluster (Figure \ref{fig:fig_3}(c) and S4, Movie S3). When we changed the confinement shape in the simulation, the inward flow still occurred, but the shape of the cluster changed according to the confinement shape (Figure S4, Movie S4). We found that simulations with \(W/D\) ratios matching the experiments and with parameters \(Pe=50\) and \(k_{\mathrm{p}}^{\mathrm{surf}}=2.5\) reproduced qualitatively similar cluster shapes and roundness trends (Figure \ref{fig:fig_3}(c), 3(d)).

Since it has been shown that the dynamics of the active fluid depend primarily on \(Pe\) and \(k_{\mathrm{p}}^{\mathrm{surf}}\) \cite{PhysRevResearch.5.013208}, we investigated whether the cluster geometry would depend on these parameters. The effect of the microwell geometry on the cluster asymmetry was more pronounced for larger \( W/D \) ratios (Figure S5), so we chose the confinement with \( W/D=0.85 \) for this purpose. We found that decreasing \(k_{\mathrm{p}}^{\mathrm{surf}}\) made the shape of the cluster rounder, but also reduced the size of the cluster, because a lower polymerization rate would deplete the amount of F-actin present in the confinement (Figure \ref{fig:fig_3}(e)). On the other hand, increasing \(Pe\) made the cluster smaller while also making it more curved, suggesting that a higher effective contractility would force F-actin into a more compact space.

To clarify the effect of \(Pe\) and \(k_{\mathrm{p}}^{\mathrm{surf}}\), it is useful to utilize the phase diagram of the shape and size changes of the cluster. The \(Pe\)-\(k_{\mathrm{p}}^{\mathrm{surf}}\) phase diagram of the roundness parameter showed it to be quite sensitive to \(k_{\mathrm{p}}^{\mathrm{surf}}\) but only slightly dependent on \(Pe\) (Figure \ref{fig:fig_3}(f)). The roundness decreased as \(k_{\mathrm{p}}^{\mathrm{surf}}\) increased, suggesting that in addition to the confinement shape, the polymerization rate also affects the contracted shape of the confined active fluid. A higher \(k_{\mathrm{p}}^{\mathrm{surf}}\) increased the ratio of the cluster to the confinement area while a larger \(Pe\) decreased it (Figure \ref{fig:fig_3}(g)). Thus, the simulations predict that the shape and size of the cluster can be regulated by the polymerization rate and the effective contractility of the actomyosin network.

\begin{figure}
\includegraphics[width=0.9\textwidth]{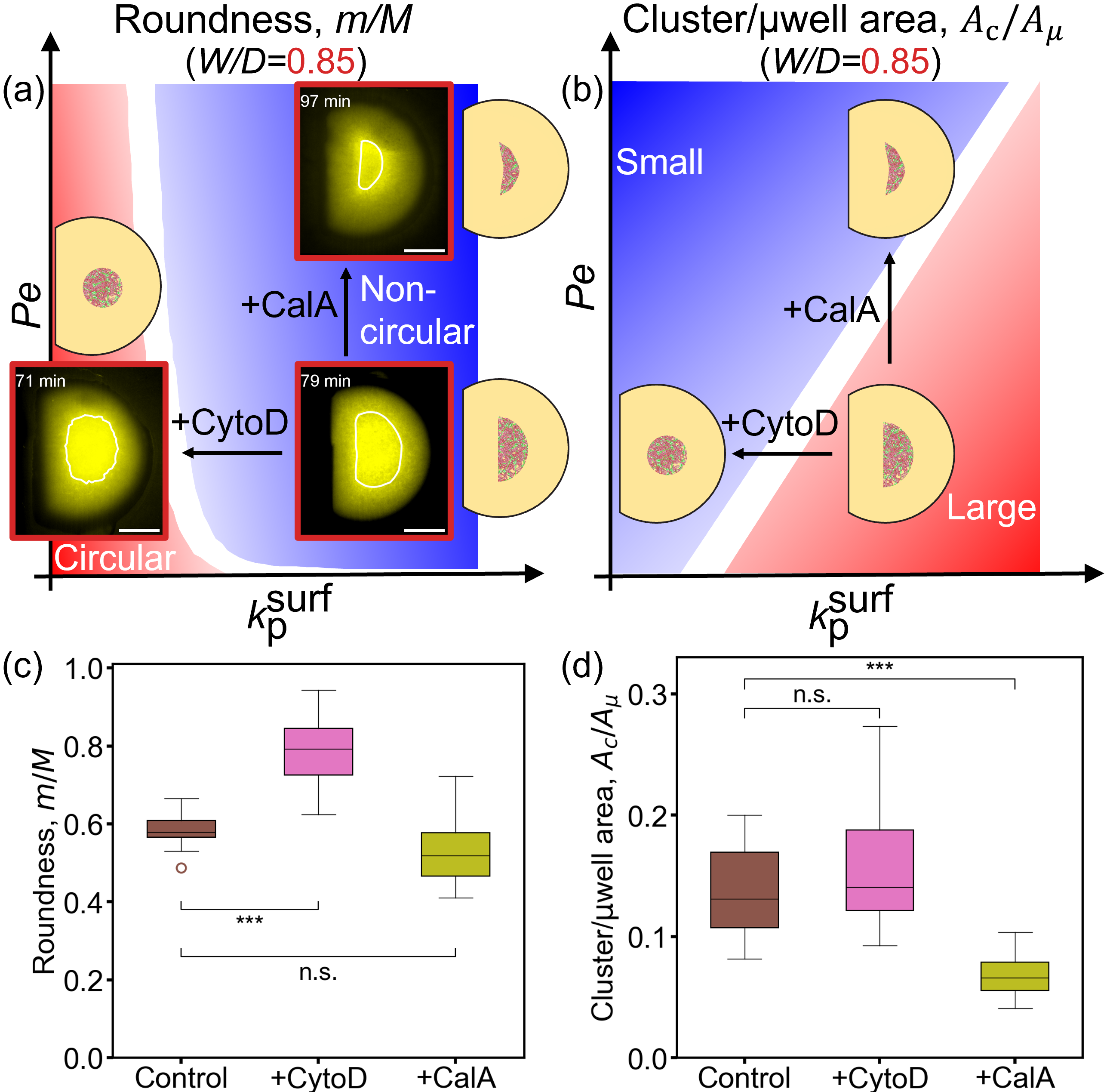}
\caption{\small{Active transport and reorganization determines the shape and size of the contracted confined actin cytoskeleton. (a) Effect of chemical perturbations on the cluster formation in microwells with \(W/D=0.85\) and \(D=350\ \upmu\)m; CytoD is an actin polymerization inhibitor while CalA is a myosin activator. The white lines denote the boundary of the clusters. Schematic diagram showing how the cluster (a) shape and (b) size in the simulations changes with \(Pe\) and \(k_{\mathrm{p}}^{\mathrm{surf}}\) and the chemical perturbations that would change these parameters in the experiments. Comparison of (c) roundness and (d) cluster-to-microwell area ratio between the control and the perturbed systems for \(W/D = 0.85\). CytoD increases cluster roundness while CalA decreases cluster area. At least \(n=8\) microwells were used for each boxplot in (c) and (d), with the microwell diameters being \(D=300, 350\) \(\upmu \)m. The \(p\)-values were calculated using the Mann-Whitney \textit{U} test, *** means \(p < 0.001\). Scale bars, 100 \(\upmu \)m.}}
\label{fig:fig_4}
\end{figure}

Based on the phase diagrams in the simulation, decreasing the polymerization rate should make the clusters rounder, while increasing the contractility should make the clusters smaller (Figure \ref{fig:fig_4}(a), 4(b)). To test this in our experiments, we performed chemical perturbations modulating actin polymerization rate and myosin contractility. In separate experiments, we added Cytochalasin D (CytoD, 1.5 \(\upmu\)M), an actin polymerization inhibitor \cite{CytoD_actin}, and Calyculin A (CalA, \(0.6\ \upmu \)M), a myosin phosphatase inhibitor \cite{CalA_myosin} that effectively increases myosin contractility to implement a decrease in \(k_{\mathrm{p}}^{\mathrm{surf}}\) and an increase in \(Pe\) respectively. Figure \ref{fig:fig_4}(a) shows the effect of these perturbations in a microwell with \(W/D=0.85\).

We again used the roundness parameter \(R\) to compare the shapes of the clusters and found that \(R\) was larger in CytoD for microwells with \(W/D=0.85\) (Figure \ref{fig:fig_4}(c)) and \(W/D \geq 0.7\) (Figure S6) compared to the control, consistent with the simulation (Figure \ref{fig:fig_3}(f)). In CalA, \(R\) did not change significantly for microwells with \(W/D=0.7,\ 0.85\) (Figure \ref{fig:fig_4}(c) and S6) and it was lower for microwells with \(W/D = 0.0,\ 0.5,\ 1.0\) (Figure S6). The minute (or negligible) reduction of \(R\) in CalA agrees with the model prediction.

The cluster size in CalA was smaller than the control for \(W/D=0.85\)  (Figure \ref{fig:fig_4}(d)) and all other microwells (Figure S6), as predicted. In CytoD, the cluster size was not smaller than the control for \(W/D=0.85\) (Figure \ref{fig:fig_4}(d)) or for any other microwell (Figure S6). This could be due to the reduction in effective actomyosin contractility upon inhibition of actin polymerization with CytoD, since the reduction in actin length could impair the force transmission within the actomyosin network. 


\begin{figure}
\includegraphics[width=0.9\textwidth]{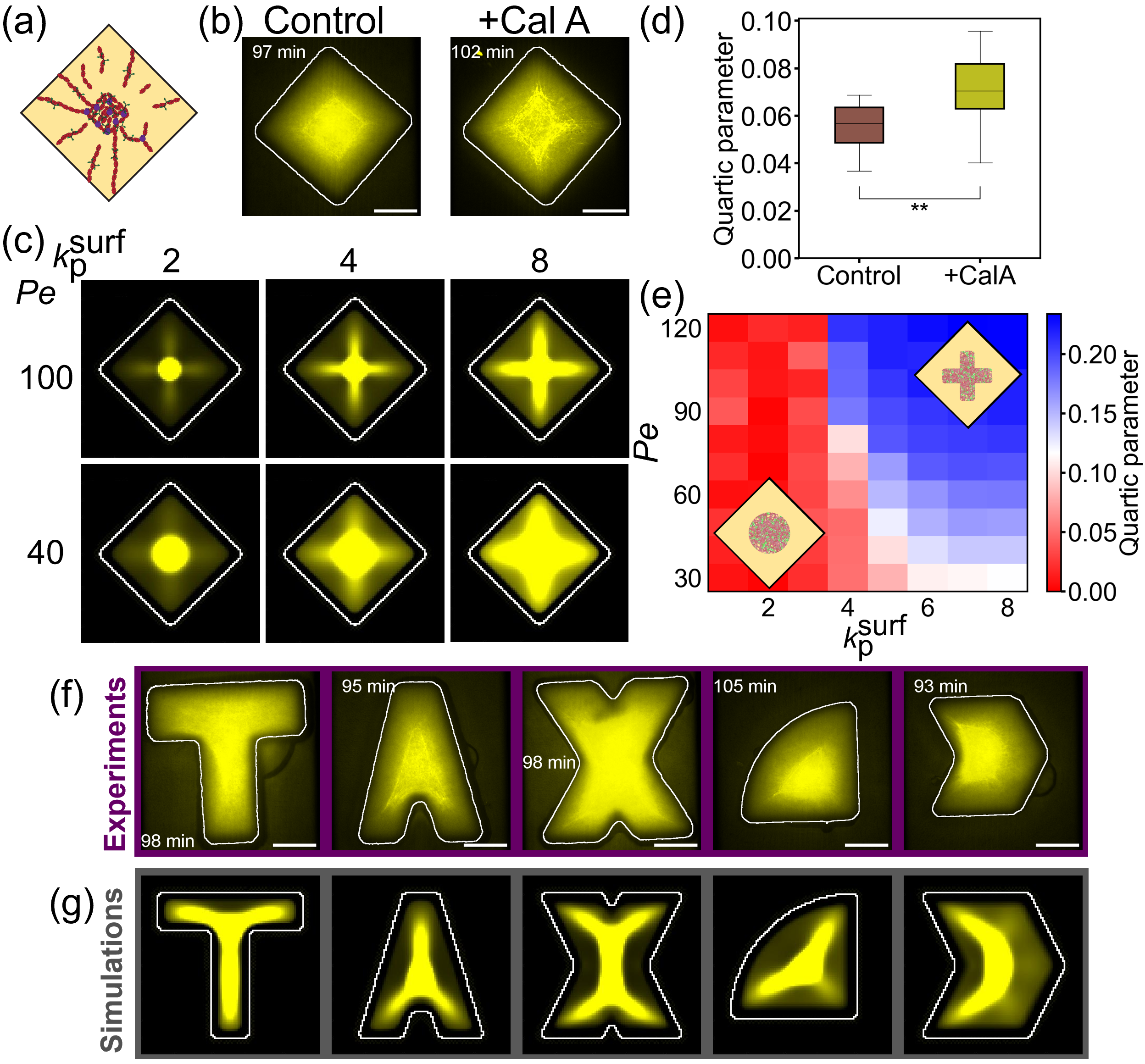}
\caption{\small{Design principles of confined actomyosin gel under spatial constraints of complex geometry. (a) Schematic of actomyosin network in a square microwell. (b) In a square microwell (side length \(=270\ \upmu\)m), the cluster shape is also that of a square (quartic parameter= 0.069); the addition of Calyculin A made the cluster shape more star-like (quartic parameter= 0.088). (c) Variation of cluster shape and size depending on \(Pe\) and \(k_{\mathrm{p}}^{\mathrm{surf}}\) in a square confinement. (d) Comparison of the quartic parameter between control (\(n=10\) microwells) and CalA (\(n=13\) microwells) perturbation in a square microwell. (e) Phase diagram plotting the dependence of the cluster shape on \(Pe\) and \(k_{\mathrm{p}}^{\mathrm{surf}}\) using a quartic parameter in a square confinement. (f) Cluster formation in complex microwells. (g) Cluster shapes in numerical simulations of the complex confinements with \(Pe=50,\ k_{\mathrm{p}}^{\mathrm{surf}}=2.5\). The \(p\)-value was calculated using the Mann-Whitney \textit{U} test, ** means \(p < 0.01\). Scale bars, 100 \(\upmu \)m.}}
\label{fig:fig_5}
\end{figure}

Next, we extend our model and experimental system to answer the question of the shape of the cluster in polygonal confinement. We confined the cell extracts inside a square microwell (Figure \ref{fig:fig_5}(a)), and found that the shape of the cluster completely reflects a square shape (Figure \ref{fig:fig_5}(b), left), suggesting that the transfer of the confinement asymmetry to the cluster is a general feature of the actomyosin network. When actomyosin was perturbed with CalA, the shape of the cluster became star-like (Figure \ref{fig:fig_5}(b), right), further suggesting that the properties of actomyosin influence its contraction. On numerically solving our model for a square confinement, we again found that the shape and size of the cluster to depend on \(Pe\) and \(k_{\mathrm{p}}^{\mathrm{surf}}\) (Figure \ref{fig:fig_5}(c)). For a low \(k_{\mathrm{p}}^{\mathrm{surf}}\), the cluster shape was circular, and increasing it turned the shape into a square. Increasing \(k_{\mathrm{p}}^{\mathrm{surf}}\) or \(Pe\) further resulted in a star-like shape. To quantify the shape, we used a generalized quartic parameter, which has larger values for shapes having a four-fold symmetry (see Materials and Methods for its definition). In the experiments, the CalA perturbation caused the quartic parameter to increase (Figure \ref{fig:fig_5}(d)). In the simulations, it increased with \(k_{\mathrm{p}}^{\mathrm{surf}}\) and also with \(Pe\) for high \(k_{\mathrm{p}}^{\mathrm{surf}}\) (Figure \ref{fig:fig_5}(e)), showing a similar trend to roundness in a semicircular confinement (Figure \ref{fig:fig_3}(f)).

Finally, we confined the cell extracts inside more complex microwells in the form of letters and irregular shapes (Figure \ref{fig:fig_5}(f)). The letter T microwell had a cluster shaped like a funnel, and the quarter circle microwell had a triangular cluster. On the other hand, in the case of letters A and X, and the arrow microwell, the cluster shape was fairly similar to the microwell shape (more examples in Figure S7). In simulations of these confinements (\(Pe=50\), \(k_{\mathrm{p}}^{\mathrm{surf}}=2.5\)), we found that the cluster shapes were qualitatively in agreement with their experimental counterparts (Figure \ref{fig:fig_5}(g)). Thus, even in complex microwells, the confinement shape can be transcribed into the cluster shape up to a certain extent, and our simulations agree well with the experiments.

\section{Discussion}
In this study, we explored the question: how does confinement geometry affect the self-organization of the actomyosin network? To address this, we confined cytoplasmic extracts containing the actomyosin network derived from \textit{Xenopus laevis} eggs within fabricated microwells. Our results showed that the asymmetry in the shape of the microwell was proportionally transferred on to the asymmetry in the shape of the contracted cluster. The clusters having a flat side was particularly interesting as one would expect a straight alignment of actin filaments to be unstable due to the contractile force of myosin and the actomyosin network to contract to form a more circular shape. This reinforces the idea that physical boundaries and confinement influence the self-organization of the cytoskeleton and, more generally, that active cytoskeletons can be controlled through geometric confinement. Another noteworthy point was the effect of sharp changes in the direction of the confinement boundaries, such as a corner in a semicircle or square. At these points, the force applied by myosin, because it is perpendicular to the surface, abruptly changes direction. This might lead to accumulation of actin and could be the reason behind the clusters having a small outgrowth at their corners\cite{schuppler2016boundaries}, which were also present in the simulations.

Moreover, we found that the morphology of the cluster was dependent on the contractile activity of myosin motors and the polymerization rate of F-actin. Altering the properties of the actomyosin network with motor-driven contraction was essential to control the size of the cluster, whereas the polymerization rate of F-actin was the primary factor that determined the cluster shape. This implies that the emergent dynamics involving molecular network reorganization, such as motor-driven advective flow, are also expected to expand the versatility for controlling hydrogel morphology. In terms of dynamics, it would be important to further develop the methodology into deformable boundary conditions such as liposomes \cite{sakamoto2024mechanical, litschel2021reconstitution, Sakamoto_combio_2024, sakamoto_prr_liposome, liu_liposome}. Exploring the effect of such interplay between deformable interface and active cytoskeletal dynamics could be a future challenge in the engineering of biocompatible hydrogels.


\section{Materials and Methods}

\subsubsection{Microscopy}
To image actomyosin dynamics, we used a confocal microscope (IX73; Olympus) with a confocal scanning unit (CSU-X1; Yokogawa Electric Cor. Ltd.), an iXon-Ultra EM-CCD camera (Andor Technologies) and a ×20 objective lens (UPlanXApo 20×/0.80; OLYMPUS) in the 561 nm fluorescence channel. For time-lapse acquisition wave formation, images were taken every 5 seconds, and for actin accumulation, Z-scans of the microwell were taken every 30 seconds. We followed the protocol from previous studies for extracting actomyosin and sample preparation \cite{Sakamoto2020,sakamoto2022geometric}.

\subsubsection{Image analysis}
Image analysis was done on Fiji and MATLAB. In short, the cluster was identified by thresholding the images. If an appropriate threshold could not be found, the clusters were manually identified. The calculation of the roundness and area of the cluster, as well as extraction of the coordinates of the cluster boundary were done using the built-in functions of Fiji. PIVlab \cite{pilvab} on MATLAB was used for PIV analysis.

The quartic parameter (Figure \ref{fig:fig_5}) is defined as follows: we perform a Fourier transform of the cluster boundary and take the ratio of the fourth to the zeroth Fourier modes. 



\subsubsection{Device microfabrication}
Detailed fabrication protocols summarized in the Supporting Information.


\section{Acknowledgements}
This work was supported by Grant-in-Aid for Transformative Research Areas (A) (JP23H04711), Grant-in-Aid for Challenging Research (Exploratory) (JP24K21534), Grant-in-Aid for Scientific Research (B) (JP23H01144), and JST FOREST Grant (JPMJFR2239).

\bibliography{main}

\end{document}